\documentclass[aps,prb,floatfix]{revtex4}
\usepackage{graphicx,bm}
\begin{document}
\title{Anatomy of Spin-Transfer Torque} 
\author{M.D. Stiles}
\affiliation{National Institute of Standards and Technology,
Gaithersburg, MD 20899-8412}
\author{A. Zangwill}
\affiliation{School of Physics, Georgia Institute of Technology,
Atlanta, GA 30332-0430}

\date{\today} 

\begin{abstract}
Spin-transfer torques occur in magnetic heterostructures because the
transverse component of a spin current that flows from a non-magnet
into a ferromagnet is absorbed at the interface.  We demonstrate this
fact explicitly using free electron models and first principles
electronic structure calculations for real material interfaces. Three
distinct processes contribute to the absorption: (1) spin-dependent
reflection and transmission; (2) rotation of reflected and transmitted
spins; and (3) spatial precession of spins in the ferromagnet.  When
summed over all Fermi surface electrons, these processes reduce the
transverse component of the transmitted and reflected spin currents to
nearly zero for most systems of interest. Therefore, to a good
approximation, the torque on the magnetization is proportional to the
transverse piece of the incoming spin current.
\end{abstract}

\maketitle 

\section{Introduction}

When a current of polarized electrons enters a ferromagnet, there is
generally a transfer of angular momentum between the propagating
electrons and the magnetization of the film. This concept of ``spin
transfer'' was proposed independently by Slonczewski
\cite{Slonczewski:1996} and Berger \cite{Berger:1996} in
1996. Experiments soon followed where anomalies in the current-voltage
characteristics of magnetic heterostructures were interpreted as
evidence for spin transfer. \cite{others} Unambiguous confirmation
came when the phenomenon of giant magnetoresistance
\cite{FertBruno:1994} was used to detect magnetization reversal in
ferromagnetic multilayers with large current densities flowing
perpendicular to the plane of the
layers. \cite{Myers:1999,Grollier:2001,Wegrowe:2001} Subsequently,
spin transfer has been implicated to explain the observation of spin
precession for high-energy, spin-polarized electrons that traverse a
magnetic thin film \cite{Weber:2001} and enhanced Gilbert damping in
magnetic multilayers compared to one-component magnetic
films. \cite{Urban:2001} More experiments may be expected in the
future because spin transfer is expected to play an important role in
the nascent field of ``spin electronics''. \cite{Prinz:1999}

Theoretical work on spin transfer generally falls into one of three
categories.  One group of articles focuses on deriving and solving
classical equations of motion for the magnetization.
\cite{Baz:2000,Sun:2000,HZE:2001,Miltat:2001,ZL:2002,Tserk:2001} These
studies generalize the Landau-Lifshitz equation to take account of
spin currents, spin accumulation, \cite{spinaccumulation} and the
mechanical torques which necessarily accompany (spin) angular momentum
transfer.  A second group of articles generalizes charge transport
theory to take account of spin currents and spin relaxation.
\cite{Weg:2000,Brataas:2001,Waintal:2000,Hernando:2000,us:2001} These
theories compute the spin-transfer torques that serve as input to the
magnetization calculations.  The torque can be computed
phenomenologically, or from the Boltzmann, Kubo or Landauer formalisms
that incorporate quantum mechanical information explicitly.  Finally,
there are articles that report quantum mechanical calculations of the
parameters that serve as input to the transport theories. The model
studies of Slonczewski \cite{Slonczewski:1996} and Berger
\cite{Berger:1996} are of this sort, as are the first-principles,
electronic structure calculations of Xia and co-workers.\cite{Xia}

In a previous paper, \cite{us:2001} the present authors used a $2 \times
2$ matrix Boltzmann equation to compute spin currents, spin
accumulation, magnetoresistance, and spin-transfer torques in a
Co/Cu/Co multilayer with non-collinear magnetization. The physics of
spin transfer entered this semi-classical, kinetic theory calculation
through quantum mechanically derived matching conditions 
imposed at each ferromagnet/non-magnet interface. Specifically, we
took account of a {\it reflection mechanism} \cite{Slonczewski:1996}
that arises because the interface reflection and transmission
amplitudes for polarized electrons are spin dependent. We also took
account of an {\it averaging mechanism} \cite{Berger:1996} that arises
because conduction electron spins precess around the magnetization
vector in each ferromagnet.
The present work was motivated originally by two assumptions we made to simplify
the Co/Cu/Co calculations. First, we set to zero the transverse
component of the spin of the conduction electron ensemble in each
ferromagnet. Second, we disregarded the phase of the reflection and
transmission amplitudes.  As best we can determine, the same
assumptions are implicit in the Landauer-type model calculations
reported in Ref.~\onlinecite{Brataas:2001} and
Ref.~\onlinecite{Waintal:2000}.  Therefore, before calculations of
this sort are carried very much further, it seemed appropriate to look
more carefully into the correctness of these assumptions.  As we will
the spin transfer process is more subtle and complex than previously
managed. 

In this paper, we analyze quantum mechanically the fate of a polarized
current that enters a ferromagnet from a metallic non-magnet.  Using
both the free electron model and first principles electronic structure
calculations, we conclude that the assumptions in question are largely
justified.  An important point is that the spin of an electron
generally rotates when it is reflected or transmitted at an
interface. This separates the reflection mechanism into two pieces. A
spin-filter effect reduces the transverse spin component of each
electron individually. A further reduction occurs when we sum over all
Fermi surface electrons because substantial phase cancellation occurs
when the distribution of spin rotation angles is broad.  As for the
mechanism we called ``averaging'' in Ref.~\cite{us:2001}, cancellation
occurs because electrons have 
different precession frequencies. This leads to an asymptotic,
oscillatory, power-law (rather than exponential) decay of the
transmitted transverse spin component. Putting everything together, we
find that (except in very exceptional cases) the transverse spin
current is almost completely absorbed within a few lattice constants
of the interface. None, or very little, is reflected or
transmitted. As a result, the spin-transfer torque is very nearly
proportional to the transverse piece of the incident spin current.

The plan of this paper is as follows. In Section II, we define the
basic variables of spin transport and establish our notation. Section
III analyzes the spin current and spin-transfer torque near a
magnetic/non-magnetic interface using a free electron model for both
materials.  Section IV generalizes the analysis of Section III to the
case of real materials. We summarize our results in Section V.

\section{Background} 
To help introduce the theory of spin transport, it is useful first to
set down the familiar equations of particle transport. These involve
the number density,
\begin{eqnarray}
n({\bf r})= \sum_{i\sigma} \psi_{i \sigma}^*({\bf r})\,\psi_{i\sigma}({\bf r}),
\label{density}
\end{eqnarray}
and the number current density,
\begin{eqnarray}
{\bf j}({\bf r}) = {\rm Re}\sum_{i \sigma}
\psi^*_{i\sigma}({\bf r})\,{\bf \hat{v}}\,\psi_{i\sigma}({\bf r}),
\label{current}
\end{eqnarray}
where ${\bf \hat{v}}=-(i\hbar / m)\nabla$ is the velocity operator and
$\psi_{i,\sigma}({\bf r})$ is an occupied single particle wave
function with state index $i$ and spin index $\sigma$. The continuity
equation,
\begin{eqnarray}
\bm\nabla \cdot {\bf j} + {\partial n \over \partial t} = 0,
\label{continuity}
\end{eqnarray}
expresses the conservation of particle number. In this paper, we will
be interested exclusively in steady-state situations where the time
derivative in (\ref{continuity}) is zero.  Not far from equilibrium,
the current takes the phenomenological form,
\begin{eqnarray}
{\bf j}= (\sigma/e) {\bf E} - D \bm\nabla \delta n,
\label{Fick}
\end{eqnarray}
where $\delta n = n - n_{\rm eq}$ is the deviation of the number
density from its equilibrium value, ${\bf E}$ is an electric field,
$\sigma$ is the conductivity, and $D$ is a diffusion constant. The
latter two are second rank tensors in the general case.

For the spin degree of freedom, the analogs to (\ref{density}) and
(\ref{current}) are the spin density,
\begin{eqnarray}
{\bf m}({\bf r})= \sum_{i \sigma\sigma'} \psi_{i\sigma}^*({\bf r})
\,{\bf s}_{\sigma,\sigma'}\,
\psi_{i\sigma'}({\bf r}),
\end{eqnarray}
and the spin current density
\begin{eqnarray}
{\bf Q}({\bf r})=
      \sum_{i \sigma\sigma'}{\rm Re}\left[
        \psi_{i\sigma}^* ({\bf r})\, 
        {\bf s}_{\sigma,\sigma'} \otimes \hat{\bf v}\,
         \psi_{i\sigma'} ({\bf r})
    \right],
\label{eq:Q}
\end{eqnarray}
where ${\bf s}= (\hbar / 2)\bm\sigma$ and $\bm\sigma$ is a vector
whose Cartesian components are the three Pauli matrices.  The spin
current is a tensor quantity. The left index of $Q_{ij}({\bf r})$ is
in spin space and the right index is in real space.  Spin is not
conserved so the analog of (\ref{continuity}) generally has non-zero
terms on the right hand side.  For our problem,
\begin{eqnarray}
\bm\nabla\cdot{\bf Q} + {\partial {\bf m} \over \partial t} =-{
\bm\delta {\bf m} \over \tau_{\uparrow\downarrow} } + {\bf n}_{\rm
ext} 
\label{spincontinuity}
\end{eqnarray}
where ${\bf n}_{\rm ext}$ is an external torque density,
$\bm\nabla\cdot{\bf Q}=\partial_k  Q_{ik}$ and
$\bm\delta {\bf m}= (|{\bf m}|-m_{\rm eq})\hat{\bf m}$ is the
so-called {\it spin accumulation}. \cite{spinaccumulation}
The first term on the right side of (\ref{spincontinuity}) accounts
for the transfer of angular momentum between the spin current and the
lattice due to spin-flip. This 
process, with relaxation time $\tau_{\uparrow\downarrow}$, changes
the magnitude of the local spin density, but not its direction.
The second term on the right side of (\ref{spincontinuity}) describes 
all external torques that act to change the direction of the local
magnetization.For example, the Landau-Lifshitz-Gilbert torque density, 
\begin{eqnarray}
{\bf n}_{\rm ext}=- (g\mu_{\rm B} / \hbar) {\bf m} \times {\bf B}_{\rm eff}
+ \alpha \,{\bf \hat{m}} \times \dot{\bf m}
\end{eqnarray}
includes an effective field ${\bf B}_{\rm eff}$  and phenomenological
damping. The effective field is due to exchange, anisotropies, and any
external fields that might be present. 

To study magnetization dynamics, we merely rearrange
(\ref{spincontinuity}) to
\begin{eqnarray}
{\partial {\bf m} \over \partial t} ={\bf n}_{\rm c} + {\bf n}_{\rm
ext}
\end{eqnarray}
where
\begin{eqnarray}
{\bf n}_{\rm c}=  -{ \bm\delta {\bf m} \over \tau_{\uparrow\downarrow} }
-\bm\nabla\cdot{\bf Q}
\label{eq:ctorque}
\end{eqnarray}
is the current-induced contribution to the torque density.
The divergence theorem
then shows that, apart from spin-flip, the torque on the total 
magnetization in a volume $V$
arises from the net flux of spin current  into and out of the surface $S$
that bounds $V$. Phenomenologically, the
spin current is driven by drift and diffusion:
\begin{eqnarray}
Q_{ik}= \sigma^{\rm p}_{i} E_k - D^{\rm p} \partial_k 
\delta m_i
\label{spinFick}
\end{eqnarray}
As in (\ref{Fick}), we assume the simplest form for the spin transport
coefficients.  That is, we use the vector $\bm\sigma^{\rm
p}={(\sigma_\uparrow-\sigma_\downarrow)}\,\hat{\bf m}$ rather than a
third rank tensor and the scalar $D^{\rm p}$ rather than a fourth rank
tensor. The conductivities $\sigma_{\uparrow}$ and
$\sigma_{\downarrow}$ refer to majority and minority electrons,
respectively.

In a non-magnet, $\sigma_{\uparrow}=\sigma_{\downarrow}$ and a spin
current arises {\em only} if there are regions of the metal where
there is a gradient in the spin accumulation, $\delta {\bf m}({\bf
r})$.  This implies that the spin density ${\bf m}({\bf r})$ and the
spin current density ${\bf Q}({\bf r})$ are only indirectly related to
each other. They need not be collinear and are generally not
proportional.  In a ferromagnet, an electric field produces a current
of polarized spins simply because
$\sigma_{\uparrow}\neq\sigma_{\downarrow}$.  This spin current is
modified by gradients in spin accumulation also. However, the second
term on the right side of (\ref{spinFick}) is valid (at most) when the
direction of the ferromagnetic magnetization is uniform in space.
Corrections are necessary when the magnetization rotates continuously
in space, {\it e.g.}, inside a domain wall.\cite{Baz:2000}

\begin{figure} 
\includegraphics{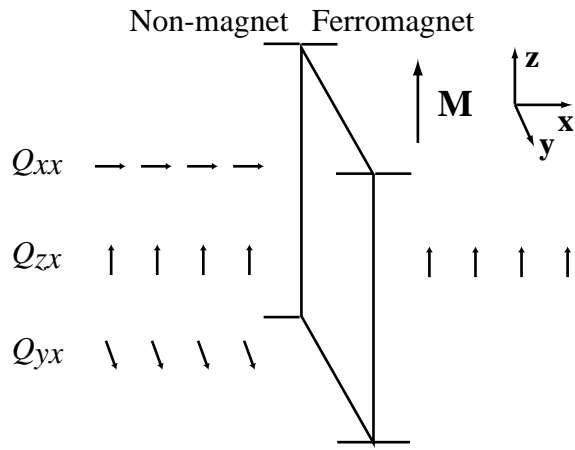} 
\caption{Three states of spin current scatter from an interface. The
current flows from left-to-right, from the non-magnet into the
ferromagnet. 
$Q_{zx}$ is longitudinal (parallel) to the  magnetization
${\bf M}$. $Q_{xx}$ and $Q_{yx}$ are transverse to ${\bf M}$. 
Only $Q_{zx}$ can be non-zero in the bulk of the magnet. The
transverse spin currents are absorbed
in the interfacial region.}
\label{fig:spin_current}
\end{figure}

With this background, the remainder of this paper is devoted to a
detailed analysis of the fate of a spin polarized current that flows
from a metallic non-magnet into a metallic, single-domain ferromagnet
through an ideal, flat interface. Specifically, we point the particle
current density vector ${\bf j}$ along positive ${\bf \hat{x}}$, we
point the ferromagnetic magnetization vector ${\bf M}$ along positive
${\bf \hat{z}}$, and we fix the interface at
$x=0$. Figure~\ref{fig:spin_current} shows three possible steady
states of pure current polarization in the non-magnet and the
associated non-zero component of the spin current density tensor. For
each case, we let only one component of $Q_{\alpha x}$ be
non-zero. $Q_{zx}\neq 0$ corresponds to longitudinal (parallel to
${\bf M}$) current polarization.  $Q_{xx}\neq 0$ or $Q_{yx}\neq 0 $
correspond to transverse (perpendicular to ${\bf M}$) current
polarization. To produce an ``incident'' polarized current in the
non-magnet, it is sufficient that the current flow into the non-magnet
from an adjacent ferromagnet and that the thickness of the non-magnet
be small compared to the non-magnet spin-flip diffusion
length. \cite{spinaccumulation} For this reason, magnetic multilayer
structures are the rule in most spin-transfer experiments.  We refer
the reader to Ref. \onlinecite{us:2001} for some insight into the
polarization process for the Co/Cu/Co system.

\begin{figure} 
\includegraphics{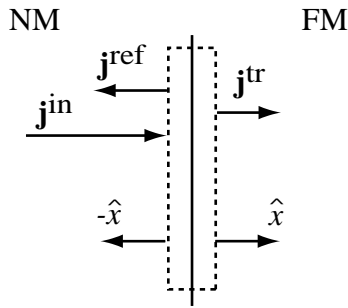} 
\caption{Interfacial pillbox used as the integration volume when
the divergence theorem is applied to (\ref{continuity}) and
(\ref{eq:ctorque}) to derive (\ref{sumj}) and (\ref{interfacetorque}).}
\label{fig:pillbox}
\end{figure}

Figure~\ref{fig:spin_current} also indicates that, of the three
incident states of pure current polarization shown, only $Q_{zx}$
transmits into the bulk of the ferromagnet.  The magnet absorbs the
transverse components. Furthermore (see below), almost none of the
transverse spin current reflects from the interface.  Therefore, if we
choose a rectangular pillbox that just straddles the interface,
the divergence theorem discussion below (\ref{eq:ctorque}) implies
that a current-induced spin-transfer torque is exerted on the
interfacial magnetization. To be more precise,
Figure~\ref{fig:pillbox} illustrates such a pillbox and incident,
reflected, and transmitted charge current density vectors.
Integrating the steady state ($\dot{\rho}=0$) version of the
continuity equation (\ref{continuity}) over the pillbox gives
\begin{equation}
0= ({\bf j}^{\rm in}-{\bf j}^{\rm tr}+{\bf j}^{\rm ref})\cdot A\hat{\bf x}
\label{sumj}
\end{equation}
where $A$ is the area of the interface.  Eq.~(\ref{sumj}) says that
the incoming flux ${\bf j}^{\rm in}\cdot A\hat{\bf x}$ minus the
outgoing flux ${\bf j}^{\rm tr}\cdot A\hat{\bf x}+{\bf j}^{\rm ref}
\cdot (-A\hat{\bf x})$ equals zero. The reflected flux has a minus
sign relative to the transmitted flux because it passes through the
opposing face of the pillbox.

Ignoring spin flip, the same integration applied to (\ref{eq:ctorque})
yields
\begin{equation}
{\bf N}_{\rm c}=\,({\bf Q}^{\rm in}-{\bf Q}^{\rm tr}
+{\bf Q}^{\rm ref})\cdot A\hat{\bf x} \simeq
\,{\bf Q}^{\rm in}_{\perp}\cdot A \hat{\bf x}
\label{interfacetorque}
\end{equation}
where ${\bf Q}^{\rm in}$, ${\bf Q}^{\rm ref}$, and ${\bf Q}^{\rm tr}$
are the spin current density (\ref{eq:Q}) computed using incident
state, reflected state, and transmitted state wave functions.
Eq.~(\ref{interfacetorque}) says that the incoming spin flux ${\bf
Q}^{\rm in} \cdot A\hat{\bf x}$ minus the outgoing spin flux ${\bf
Q}^{\rm tr} \cdot A\hat{\bf x}+{\bf Q}^{\rm ref} \cdot (-A\hat{\bf
x})$ equals the torque on the magnetization inside the
pillbox. \cite{sign_convention} The torque ${\bf N}_{\rm c}$ is a
vector in spin space because we have contracted the space index of the
spin current density with the space vector $\hat{\bf x}$.  The
approximate form on the right of Eq.~(\ref{interfacetorque}) says that
the torque is proportional to the tranverse part of ${\bf Q}_{\rm in}$.
That is the
main message of this paper. The following sections are devoted to a
demonstration that the tranverse transmitted and reflected spin currents do
indeed disappear in the immediate vicinity of the interface.

\section{Free electrons}

In this section, we compute the spin current near the interface of a
non-magnet and a ferromagnet assuming that a free electron description
is adequate for the conduction electrons in the non-magnet and also for
both the majority and minority conduction electrons in the
ferromagnet. We do this in the interest of analytic simplicity and
also because some authors \cite{HZE:2001,ZL:2002} believe this is a
fair representation of reality for the purposes of transport
calculations.

We first work out the problem of one electron scattering from a planar
interface to determine the amplitudes for reflection and
transmission. They turn out to be spin dependent.  As first shown by
Slonczewski,\cite{Slonczewski:1996} this fact alone generates a
``spin-filter'' torque because the wave function for an incident
electron with a non-zero component of spin transverse to ${\bf M}$ can
always be re-expressed in terms of up and down spin components.

The actual current polarization in the metal is obtained by summing
over the full distribution of conduction electrons. This introduces
two effects. The first arises because the reflection amplitude for
free electron interface scattering is complex. This means that the
spin of an incoming electron rotates upon reflection.  The
cancellation which occurs when we sum over all these spin vectors
reduces the net transverse spin current because reflection and
transmission both contribute to the outgoing flux from the interface
region. A second effect arises because up and down spin electrons on
the Fermi surface with the same wave vector in the non-magnet no
longer have the same wave vector when they transmit into the
ferromagnet. The two states are coherent, so precession in space
(rather than time) occurs. The precession frequency is different for
electrons from different portions of the Fermi surface.  Therefore,
when we sum over all conduction electrons, almost complete
cancellation of the transverse spin occurs after propagation into the
ferromagnet by a few lattice constants.

\subsection{Spin currents for a single electron}

Let us choose the spin quantization axis to be parallel to the
magnetization of the ferromagnet.  Then, in the non-magnet, the wave
function for an electron whose spin points in an arbitrary direction
can always be written as a linear combination of spin up and spin down
components. Specifically,
\begin{eqnarray}
\psi_{\rm in}= \left[
   \cos\textstyle{1\over2}\theta\,e^{-i\phi/2} \left| \uparrow \right\rangle
+  \sin\textstyle{1\over2}\theta\,e^{ i\phi/2} \left| \downarrow \right\rangle
\right]
e^{ik_x x} e^{i{\bf q}\cdot{\bf R}}
\label{incidentstate}
\end{eqnarray}
represents a free electron propagating toward the interface in
Figure~\ref{fig:spin_current} with its spin pointed in the direction
($\theta,\phi$) with respect to ${\bf M}$. We are interested in
conduction electrons so the wave vector ${\bf k}=(k_x, {\bf q})$
satisfies $\hbar k^2/2m=E_{\rm F}$. The spatial variable is ${\bf
r}=(x,{\bf R})$.  As the notation indicates, (\ref{incidentstate}) is
the incident state for a scattering problem that determines the wave
function for the entire system. The latter describes a steady-state
situation like current flow.\cite{Gottfried} Like the incident state
(\ref{incidentstate}), the complete scattering state can also be
written as a linear combination of spin up and spin down components:
\begin{equation}
\psi=\psi_\uparrow + \psi_\downarrow.
\label{totalpsi}
\end{equation}
In detail, 
\begin{eqnarray}
\psi_\uparrow =\cos\textstyle{1\over2}\theta\,e^{-i\phi/2}
\left|\uparrow\right\rangle \cases{ 
( e^{ik_x x} + R_{\uparrow} e^{-ik_x x} ) e^{i{\bf q}\cdot{\bf R}}&
$~~x<0$\cr 
T_{\uparrow} e^{ik_x^\uparrow x} e^{i{\bf q}\cdot{\bf R}}&
$~~x>0$\cr} 
\nonumber\\
\psi_\downarrow =\sin\textstyle{1\over2}\theta\,e^{ i\phi/2}
\left|\downarrow\right\rangle \cases{ 
( e^{ik_x x} + R_{\downarrow} e^{-ik_x x} ) e^{i{\bf q}\cdot{\bf R}}&
$~~x<0$\cr
T_{\downarrow} e^{ik_x^\downarrow x} e^{i{\bf q}\cdot{\bf R}}&
$~~x>0$\cr} \nonumber \\ 
\label{eq:eigen}
\end{eqnarray}
 \\
where $R_{\uparrow}$, $R_{\downarrow}$, $T_{\uparrow}$, and
$T_{\downarrow}$ are the reflection and transmission amplitudes for up
and down spin electrons.  These amplitudes do not depend on the angles
$\theta$ and $\phi$.  Notice that the up and down spin components do
not propagate with the same wave vector for $x>0$.  
The wave vectors differ because their kinetic energy depends on the
exchange potential energy in the ferromagnet. 
The common
factor of $\exp(i{\bf q}\cdot{\bf R})$ in (\ref{eq:eigen}) reminds us
that scattering from a flat interface conserves the wave vector
component parallel to the interface.

\begin{figure} 
\includegraphics[height=6in]{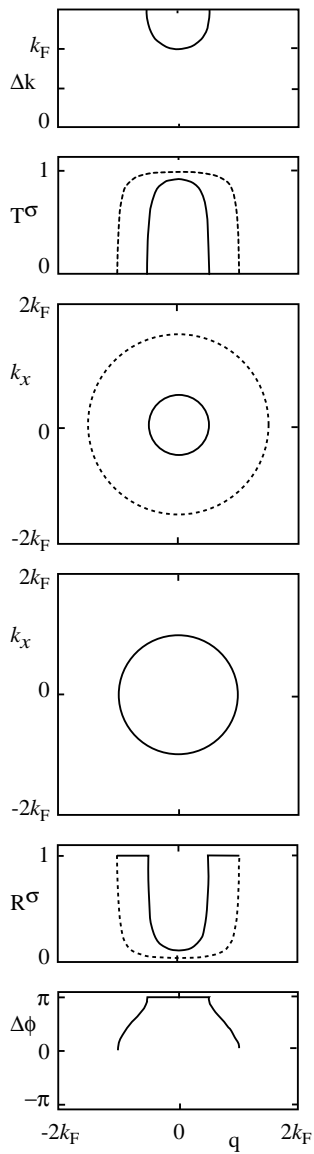} 
\caption{Slices through a set of free electron Fermi surfaces. The two
middle panels show the Fermi surface for the non-magnet and the
superimposed Fermi surfaces of the
majority (dashed) and minority (solid) states of the ferromagnet.
The panel just above the magnetic
Fermi surfaces is the probability for transmission into the ferromagnet
for  majority (dashed) and minority (solid) electrons. The panel just below
the non-magnetic Fermi surface is the probability for reflection back into the
non-magnet for majority (dashed) and minority (solid) electrons. The bottom
panel shows the phase in (\protect\ref{eq:deltaphi}) acquired
by an electron because its spin rotates upon
reflection. The top panel shows the wave vector difference (\protect\ref{eq:Deltak})
for a transmitted electron.}
\label{fig:fe_slice}
\end{figure}

The transmission and reflection amplitudes are determined by the
magnitude of the potential step at the interface. For a constant
effective mass, this step height is parameterized by $k_{\rm F} ,
k_{\rm F}^\uparrow$, and $k_{\rm F}^\downarrow < k_{\rm F}^\uparrow$,
the Fermi wave vectors for, respectively, electrons in the non-magnet,
majority electrons in the ferromagnet, and minority electrons in the
ferromagnet.  The usual quantum mechanical matching conditions yield
the {\em real} transmission amplitudes
\begin{eqnarray}
T_\sigma (q)= { 2 k_x(q) \over 
                  k_x(q) + k^\sigma_x(q) }\
\label{T}
\end{eqnarray}
\\
where $k_x(q) = \sqrt{ k_{\rm F}^2 - q^2 }$ and
$k^\sigma_x(q) = \sqrt{ (k^\sigma_{\rm F})^2 - q^2 }$.
The reflection amplitudes are real or
complex depending on the magnitude of the parallel wave vector. They are
\\
\begin{equation}
R_\sigma (q)= { k_x(q) - k^\sigma_x(q) \over 
                  k_x(q) + k^\sigma_x(q) } \hskip25pt {\rm if } \hskip25pt
q^2 \le (k^\sigma_{\rm F})^2
\label{real}
\end{equation}
and
\begin{equation}
R_\sigma (q)= { k_x(q) - i\kappa^\sigma_x(q) \over 
                  k_x(q) + i\kappa^\sigma_x(q) } \hskip25pt {\rm if} \hskip25pt
q^2 > (k^\sigma_{\rm F})^2
\label{imag}
\end{equation}
\\ where $\kappa^\sigma_x(q) = \sqrt{ q^2 - {k^\sigma_{\rm F}}^2}$.
The associated transmission and reflection {\em probabilities},
\begin{eqnarray}
{\rm R}^\sigma(q)&=& | R_\sigma (q) |^2 \nonumber\\
{\rm T}^\sigma(q)&=& { k^\sigma_x(q) \over k_x(q) }| T_\sigma (q) |^2,
\label{probs}
\end{eqnarray}
satisfy ${\rm R}^\sigma+{\rm T}^\sigma=1$ and are plotted in
Figure~\ref{fig:fe_slice} for a slice through the free electron Fermi
surfaces defined by $k_{\rm F}^\uparrow /k_{\rm F}=1.5$ and $k_{\rm
F}^\downarrow / k_{\rm F}=0.5$.  For this case, the transmission
probability for majority electrons (dashed curve) is unity near the
zone center and then falls rapidly to zero near $k_{\rm F}$.  The
minority electrons (solid curve) transmit similarly except that ${\rm
T}^{\downarrow}$ falls to zero near $k_{\rm F}^\downarrow$.

It is now straightforward to compute and interpret the incident,
reflected, and transmitted number current densities and spin current
densities. We need only (\ref{current}), (\ref{eq:Q}), and the
appropriate piece of the wave function (\ref{eq:eigen}). The incident
current densities are
\begin{eqnarray}
j^{\rm in}_x&=& v_x 
\nonumber\\
Q^{\rm in}_{xx} &=&
 {\hbar \over 2}  
 v_x
\sin\theta \cos\phi
\nonumber\\
Q^{\rm in}_{yx} &=&
 {\hbar \over 2}  
 v_x
\sin\theta \sin\phi
\nonumber\\
Q^{\rm in}_{zx}&=&
 {\hbar \over 2}  
 v_x
\cos\theta
\end{eqnarray}
where $v_x =\hbar k_x / m$. The reflected current densities are 
\begin{eqnarray}
j^{\rm ref}_x &=& -|v_x| 
\left[ \cos^2\textstyle{1\over2}\theta\,|R_\uparrow|^2  
     + \sin^2\textstyle{1\over2}\theta\,|R_\downarrow|^2 \right]
\nonumber\\
Q^{\rm ref}_{zx} &=& -{\hbar \over 2} |v_x| 
\left[\cos^2\textstyle{1\over2}\theta\,|R_\uparrow|^2  
     - \sin^2\textstyle{1\over2}\theta\,|R_\downarrow|^2 \right]
\nonumber\\
Q^{\rm ref}_{xx}&=& -{\hbar \over 4} |v_x|
\sin\theta\, 
{\rm Re}\left[ R_\uparrow^* R_\downarrow e^{-i\phi} \right]
\nonumber\\
Q^{\rm ref}_{yx}&=& -{\hbar \over 4} |v_x|
\sin\theta\,
{\rm Im}\left[ R_\uparrow^* R_\downarrow e^{-i\phi} \right].
\label{reflect}
\end{eqnarray}

The transmitted current densities are
\begin{eqnarray}
j^{\rm tr}_x &=& 
  v_x^\uparrow\cos^2\textstyle{1\over2}\theta\,|T_\uparrow|^2  
+ v_x^\downarrow\sin^2\textstyle{1\over2}\theta\,|T_\downarrow|^2 
\nonumber\\
Q^{\rm tr}_{zx} &=& 
  {\hbar \over 2} v_x^\uparrow \cos^2{\textstyle{1\over2}}\theta\,|T_\uparrow|^2  
- {\hbar \over 2} v_x^\downarrow \sin^2{\textstyle{1\over2}}\theta\,|T_\downarrow|^2 
\nonumber\\
Q^{\rm tr}_{xx}({\bf r}) &=&{\hbar \over 4} 
{v_x^\uparrow+v_x^\downarrow\over 2 }
\sin\theta\,
{\rm Re}\left[ T_\uparrow^* T_\downarrow e^{-i\phi} 
e^{i(k_x^\downarrow-k_x^\uparrow)x} \right]
\nonumber\\
Q^{\rm tr}_{yx}({\bf r}) &=& {\hbar \over 4} 
{v_x^\uparrow+v_x^\downarrow\over 2 } 
\sin\theta\,
{\rm Im}\left[ T_\uparrow^* T_\downarrow e^{-i\phi} 
e^{i(k_x^\downarrow-k_x^\uparrow)x} \right]
\nonumber\\
\label{transmit}
\end{eqnarray}
where $v_x^\sigma=\hbar k_x^\sigma/m$.  Using (\ref{probs}), it is
easy to check that $j^{\rm in}_x=j^{\rm tr}_x - j^{\rm ref}_x$ and
$Q^{\rm in}_{zx}=Q^{\rm tr}_{zx}-Q^{\rm ref}_{zx}$.  The first
relation is consistent with (\ref{sumj}) because there is no
accumulation of charge at the interface. Using
(\ref{interfacetorque}), the second relation tells us that there is no
torque associated with the transport of longitudinal spin current.
However, a similar relationship does {\em not} hold for the other two
components of ${\bf Q}$. There is a discontinuity in the transverse
spin current when a spin scatters from an interface.  According to
(\ref{interfacetorque}), this implies that a current-induced torque
acts on the magnetization. In fact, three distinct mechanisms
contribute to the net torque.

One source of discontinuity and spin-transfer torque is {\em spin
filtering}. This occurs when the reflection probabilities are spin
dependent.\cite{Slonczewski:1996} To see this, note first that the
specific superposition of up and down spin components displayed in the
incident state wave function (\ref{incidentstate}) corresponds to a
specific transverse component of the spin vector.  If $R_\uparrow =
R_\downarrow$ and $T_\uparrow=T_\downarrow$, that specific linear
combination is preserved in the reflected and transmitted pieces of
the scattering state and no discontinuity occurs in the spin current.
However, if the reflection and transmission amplitudes {\it differ}
for up and down spin components, the up and down spin content of the
spatially separated reflected and transmitted states differ from one
another.  This leads unavoidably to different transverse spin
components and thus to a discontinuity in the transverse spin current.
Given the structure of (\ref{reflect}) and (\ref{transmit}), we use
the reflection and transmission {\em probabilities} in the combination
$\sqrt{ {\rm R}^\uparrow {\rm R}^\downarrow} + \sqrt{ {\rm T}^\uparrow
{\rm T}^\downarrow}$ as a measure of the ability of spin filtering to
provide spin-transfer torque. The next-to-top and next-to-bottom
panels in Figure~\ref{fig:fe_slice} display the required information.

A second source of transverse spin current discontinuity and
spin-transfer torque is {\em spin rotation}. This occurs when the
product $R^*_\uparrow R_\downarrow$ is not positive
real. Specifically, (\ref{reflect}) shows that the transverse
components of the reflected spin current contain a factor
\begin{eqnarray}
R_\uparrow^* R_\downarrow = | R_\uparrow^* R_\downarrow |
e^{i\Delta\phi}.
\label{eq:deltaphi}
\end{eqnarray}
The phase $\Delta\phi$ evidently adds directly to the azimuthal angle
$\phi$ used to define the spin direction in the incident state vector
(\ref{incidentstate}).  In other words, the reflected spin direction
rotates with respect to the incident spin direction. This is an
entirely quantum mechanical phenomenon for which there is no classical
analog.  The bottom panel of Figure~\ref{fig:fe_slice} shows that the
range of $\Delta\phi$ can be surprisingly large. Indeed, for this
choice of Fermi surfaces, the spin direction completely reverses when
an electron reflects from the interface at near-normal
incidence. There is no corresponding rotation for transmitted
electrons because $T_{\uparrow}$ and $T_{\downarrow}$ are positive
real (for free electrons).  The resulting discontinuity in the
transverse spin current leads to a spin-transfer torque that is
distinct from spin filtering.

Finally, a glance at (\ref{transmit}) reveals that {\em spin
precession} is a third source of spin-transfer torque. Note specially
the spatially-varying phase factors which appear in the transmitted
transverse spin currents because $k_x^\uparrow \neq k_x^\downarrow$ in
the ferromagnet.  Their net effect is spatial precession because
$Q_{xx}$ and $Q_{yx}$ simply rotate into one another as a function of
$x$.\cite{precess} From (\ref{eq:ctorque}), such a spatial variation
of ${\bf Q}$ contributes a distributed torque density at every point
in the ferromagnet.  The top panel of Figure~\ref{fig:fe_slice} shows
the range of spatial precession ``frequencies''
\begin{equation}
\Delta {\rm
k}=k_x^\downarrow-k_x^\uparrow
\label{eq:Deltak}
\end{equation}
for the free electron model of that
figure.

\subsection{Spin currents for a distribution of electrons}
\label{sec:free_dist}

The spin currents relevant to experiment reflect the combined effect
of all the conduction electrons. In the most general description of
transport, it is
necessary to keep track of the quantum mechanical coherence between all
electrons in different eigenstates. However, to model the
spin-transfer torque experiments reported to date,
\cite{Myers:1999,Grollier:2001} it is not necessary to maintain the
coherence between states with different Fermi surface wave vectors.
It is sufficient to use a semi-classical theory that maintains only the
coherence between up and down spin states at each ${\bf k}$-point on
the Fermi surface. Accordingly, we define a $2 \times 2$ electron
occupancy distribution matrix
\begin{eqnarray}
{\bf f}({\bf k},{\bf r}) = U({\bf k},{\bf r})  
\left( \begin{array}{cc} f_\uparrow({\bf k},{\bf r}) & 0 \\ 0
&f_\downarrow({\bf k},{\bf r}) \end{array}\right) U^\dagger({\bf
k},{\bf r})
\label{eq:matrix}
\end{eqnarray}
in terms of the scalar occupancy functions
for up and down spins and the spinor rotation matrix
\begin{eqnarray}
U({\bf k},{\bf r}) =\left( \begin{array}{cc}
\cos(\theta/2) e^{-i\phi/2}  & -\sin(\theta/2) e^{-i\phi/2} \\
\sin(\theta/2) e^{ i\phi/2}  &  \cos(\theta/2) e^{ i\phi/2}\end{array}
\right). 
\label{eq:unitary}
\end{eqnarray}
We have suppressed the ${\bf k}$ and ${\bf r}$ dependence of $\theta$
and $\phi$ for simplicity.

Elsewhere, we have solved the Boltzmann equation to find {\bf f}({\bf
k},{\bf r}) for a typical spin-transfer geometry.\cite{us:2001}  For
the simple scattering problem treated here, the reflected and
transmitted distributions are determined entirely by the reflection
and transmission amplitudes and the incident electron distribution at
the interface between the non-magnet and the ferromagnet: ${\bf
f}({\bf k})={\bf f}(x=0,{\bf k})$.  For this distribution, the
semi-classical version of the spin current (\ref{eq:Q}) is
\begin{eqnarray}
{\bf Q}^{\rm in}&=& {\hbar \over 2 }\int\limits_{v_{x}>0} { d^3 k
\over (2\pi)^3 } 
{\rm Tr}\left[ {\bf f({\bf k})} \bm\sigma \right] \otimes {\bf v}({\bf k}).
\label{Qin}
\end{eqnarray}
The restriction ${v_{x}>0}$ limits the integration to the occupied
electron states that move toward the interface. Using
(\ref{eq:matrix}), (\ref{eq:unitary}) and the cyclic properties of the
trace, we get, {\em e.g.},
\begin{eqnarray}
{Q}^{\rm in}_{xx}&=&  {\hbar \over2}\int\limits_{v_{x}>0} { d^3 k
\over (2\pi)^3 } f_{\rm p}({\bf k})v_{x}({\bf k}) \sin\theta_{\bf k}
\cos\phi_{\bf k} ,
\label{Qxxin}
\end{eqnarray}
where $f_{\rm p}({\bf k}) =f_\uparrow({\bf k}) - f_\downarrow({\bf
k})$ determines the degree of polarization at each point on the
Fermi surface. The angles $\theta_{\bf k}$ and $\phi_{\bf k}$ determine
the direction of the spin polarization.  Electron states in the immediate
vicinity of the Fermi surface dominate the transport of charge and
spin.  Therefore, we write
\begin{eqnarray}
f_\sigma({\bf k}) \rightarrow f_0(\epsilon_{\bf k}) + g_\sigma({\bf q})
{\partial f_0 (\epsilon_{\bf k} ) \over \partial\epsilon_{\bf k}}
\label{linearize}
\end{eqnarray}
where $f_0$ is the equilibrium Fermi-Dirac distribution function and
the partial derivative restricts ${\bf k}$ to the Fermi surface. We
write $g_\sigma({\bf q})$ rather than $g_\sigma({\bf k})$ because
$|{\bf k}|^2=k_x^2 + q^2 = k_{\rm F}^2$.  The equilibrium term does
not contribute to the spin current. Otherwise, we let
$d^3k=d^2q\,dk_x$ and use $\int dk_x \partial f_0 / \partial
\epsilon_{\bf k}=1/\hbar |v_x({\bf q})|$ in (\ref{Qxxin}). The result
is
\begin{eqnarray}
Q^{\rm in}_{xx}&=&{1 \over 4\pi } \int\limits_{v_{x}>0}  { d^2 q \over (2\pi)^2 } 
g_{\rm p}({\bf q})
\sin\theta_{\bf q} \cos\phi_{\bf q}
\label{qinxx}
\end{eqnarray}
where
\begin{eqnarray}
g_{\rm p}({\bf q}) =
g_\uparrow({\bf q}) - g_\downarrow({\bf q}).
\label{gp}
\end{eqnarray}
For $Q^{\rm in}_{yx}$, change $\cos\phi_{\bf q}$ to $\sin\phi_{\bf q}$
in (\ref{qinxx}).


The reflected spin current due to all the conduction electrons is
\begin{eqnarray}
{\bf Q}^{\rm ref}({\bf r})&=& {\hbar \over 2 }
\int\limits_{v_{x}>0} { d^3 k \over (2\pi)^3 } 
{\rm Tr}\left[ {\bf R}^\dagger({\bf k},{\bf r}) 
{\bf f({\bf k})} {\bf R}({\bf k},{\bf r}) \bm\sigma \right]  \otimes
{\bf v}^{\rm ref}({\bf k}) 
\label{Qcancel}
\end{eqnarray}
where 
\begin{eqnarray}
{\bf R}({\bf k}) = \left(\begin{array}{cc}
R_\uparrow({\bf k})e^{i{\bf k}\cdot{\bf r}}   & 0 \\
0 &R_\downarrow({\bf k})e^{i{\bf k}\cdot{\bf r}}   \end{array}\right)
\label{rmatrix}
\end{eqnarray}
and ${\bf v}^{\rm ref}({\bf k})$ is the velocity of a reflected
electron with wave vector ${\bf k}$. The ${\bf r}$-dependent
phase factors in (\ref{rmatrix}) cancel out in (\ref{Qcancel}) so,
{\em e.g.},
\begin{eqnarray}
Q^{\rm ref}_{xx}&=&  -{1 \over 4\pi } 
\int\limits_{v_{x}>0}  { d^2 q \over (2\pi)^2 } 
g_{\rm p}({\bf q})
\sin\theta_{\bf q} 
\left| R_{\uparrow}^*({\bf q}) R_{\downarrow}({\bf q})\right|
{\rm Re}\left[  e^{-i(\phi_{\bf q}-\Delta\phi_{\bf q})} 
 \right]
\label{qxxr}
\end{eqnarray}
where $\Delta\phi_{\bf q}$ is the relative phase of the
reflection amplitude as in (\ref{eq:deltaphi}).  For $Q_{yx}^{\rm
ref}$, change Re to Im in (\ref{qxxr}).

Finally, the total
transmitted spin current is
\begin{eqnarray}
{\bf Q}^{\rm tr}({\bf r})&=& {\hbar \over 2 }
\int\limits_{v_{x}>0} { d^3 k \over (2\pi)^3 } 
{\rm Tr}\left[ {\bf T}^\dagger({\bf k},{\bf r}) 
{\bf f({\bf k})} {\bf T}({\bf k},{\bf r}) \bm\sigma \right] \otimes
{\bf v}^{\rm tr}({\bf k})  
\label{Qtr}
\end{eqnarray}
where
\begin{eqnarray}
{\bf T}({\bf k},{\bf r}) = \left(\begin{array}{cc} T_\uparrow({\bf k})
e^{i{\bf k}^\uparrow\cdot{\bf r}} & 0 \\ 0 &T_\downarrow({\bf k})
e^{i{\bf k}^\downarrow\cdot{\bf r}} \end{array}\right) 
\label{eq:T}
\end{eqnarray}
and
\begin{eqnarray}
{\bf v}^{\rm tr}({\bf k}) =
{ {\bf v}^{\uparrow}({\bf k}) + {\bf v}^{\downarrow}({\bf k}) \over 2 }.
\label{avg}
\end{eqnarray}
In these formulae, the wave vector for incident states, ${\bf k}$,
transforms to either ${\bf k}^{\uparrow}$ or ${\bf k}^{\downarrow}$
when the electron enters the ferromagnet. The average transmitted
velocity ${\bf v}^{\rm tr}({\bf k})$ is defined only at values of
${\bf q}$ where both spins transmit.  A comparison of, say,
\begin{eqnarray}
Q^{\rm tr}_{xx}(x) &=&  {1 \over 4\pi } \int\limits_{v_{x}>0}  { d^2 q
\over (2\pi)^2 } g_{\rm p}({\bf q})\sin\theta_{\bf q} {
v^{\uparrow}_x({\bf q}) + v^{\downarrow}_x({\bf q}) \over | 2 v_x({\bf
q}) | }{\rm Re}\left[ T_{\uparrow}^*({\bf q})   T_{\downarrow}({\bf
q})e^{-i\phi_{\bf q}} e^{-i ( k_x^{ \uparrow} - k_x^{ \downarrow} ) x
}\right] 
\label{qxxtr}
\end{eqnarray}
with (\ref{transmit}) confirms that (\ref{Qtr}) is correct with the
definitions ({\ref{eq:T}) and (\ref{avg}).  For $Q_{yx}^{\rm tr}$,
change Re to Im in (\ref{qxxtr}).  

At this point, we must make a
specific choice for $g_{\rm p}({\bf k})$ and the polarization of the
incident spin current.  Let us assume the current is completely spin
polarized along $+\hat{\bf x}$.  This fixes $\theta_{\bf k}=\pi/2$ and
$\phi_{\bf k}= 0$.  For the distribution (\ref{gp}), we begin with the
approximate form
\begin{eqnarray}
g_{\rm p}({\bf q}) = a+ b v_x({\bf q}).
\label{distribution}
\end{eqnarray}
The two terms account for interface and bulk effects, respectively.
The velocity-dependent bulk term is familiar from textbook treatments
of electrical conductivity \cite{Ziman} except, from (\ref{spinFick}),
gradients in spin accumulation (rather than electric potential) drive
the spin current in the non-magnet.  The constant term is needed
because a spin-dependent chemical potential difference $\Delta \mu$
across an interface also drives a spin
current. \cite{spinaccumulation} In this paper, we assume that the
interface resistance is large (large reflection probability) so we use
\begin{equation}
g_{\rm p}({\bf q}) \simeq a = \Delta \mu.
\end{equation}
This is the same approximation that is made in Landauer-type
transport calculations.

With these choices, the incident spin current is  
\begin{equation}
Q_{xx}^{\rm in}=  {1\over 2} { 1 \over (2\pi)^2}
\int\limits_0^{k_{{\rm F}}} d q\, q  \Delta\mu= {1\over 4} { k_{\rm
F}^2 \over (2\pi)^2} { \Delta\mu  }.
\end{equation}
The reflected spin currents normalized to the incident spin current are
\begin{eqnarray}
{ Q_{xx}^{\rm ref} \over Q_{xx}^{\rm in} }
&=& - { 2\over k_{\rm F}^2}
\int\limits_0^{k_{\rm F}} d q\, q 
|{R_\uparrow^*(q)}
R_{\downarrow}(q)|\cos \Delta \phi_{\rm q}
\label{Qxxcos}
\end{eqnarray}
and
\begin{eqnarray}
{ Q_{yx}^{\rm ref} \over Q_{xx}^{\rm in} }
&=&  -{ 2 \over k_{\rm F}^2}
\int\limits_0^{k_{\rm F}} d q\, q 
|{R_\uparrow^*(q)}
R_{\downarrow}(q)|\sin \Delta \phi_{\rm q}.\label{Qyxsin}
\end{eqnarray}
We get $Q_{yx}^{\rm ref} \neq 0$ because, as discussed earlier, many
of the spins rotate upon reflection. On the other hand, the
sinusoidal factors lead to substantial self-cancellation of the
integrals (\ref{Qxxcos}) and (\ref{Qyxsin}) when the range of $\Delta
\phi_{\rm q}$ is large (see bottom panel of
Figure~\ref{fig:fe_slice}).\cite{Xia} In most cases, we find the
total transverse reflected spin current to be very small.

The
normalized transmitted spin currents are
\begin{eqnarray}
{ Q_{xx}^{\rm tr} (x) \over Q_{xx}^{\rm in} }&=&  { 2 \over k_{\rm
F}^2}\int\limits_0^{k_{\rm F}^\downarrow} d q\, q {
k^{\downarrow}_x(q) + k^{\uparrow}_x(q) \over  2 | k_x(q) |
}T_\uparrow(q)T_{\downarrow}(q)\cos\left[ ( k^{\downarrow}_x(q) -
k^{\uparrow}_x(q) ) x\right] 
\label{xdecay}
\end{eqnarray}
and
\begin{eqnarray}
{ Q_{yx}^{\rm tr} (x) \over Q_{xx}^{\rm in} }
&=&  { 2 \over k_{\rm F}^2}
\int\limits_0^{k_{\rm F}^\downarrow} d q\, q 
{ k^{\downarrow}_x(q) + k^{\uparrow}_x(q) \over  2 | k_x(q) | }
T_\uparrow(q)
T_{\downarrow}(q)
\sin\left[ ( k^{\downarrow}_x(q) - k^{\uparrow}_x(q) ) x\right].
\label{ydecay}
\end{eqnarray}
Based on the behavior of the transverse reflected spin current, we expect 
(\ref{xdecay}) and (\ref{ydecay}) to decay as a function of $x$
because the generally wide range of
$\Delta k= k^{\downarrow}_x(q) - k^{\uparrow}_x(q)$
(see top panel of
Figure~\ref{fig:fe_slice}) ought to induce   
self-cancellation of the integrals. In fact, like a similar
integral that appears in the theory of oscillatory exchange
coupling, \cite{Stilesintegral} we can extract the 
asymptotic form ($x \rightarrow \infty$) analytically
using a stationary phase
approximation. Only small values of $q$
contribute in that instance so for, say, the $xx$ component, we find
\begin{eqnarray}
\lim_{x\rightarrow\infty}
{ Q_{xx}^{\rm tr} (x) \over Q_{xx}^{in} }
&=& - 2 { k_{\rm F}^\uparrow  k_{\rm F}^\downarrow \over k_{\rm F}^2
}{ k_{\rm F}^\downarrow + k_{\rm F}^\uparrow \over  2 k_{\rm F}
}T_\uparrow(0)T_{\downarrow}(0){ \sin\left[ ( k_{\rm F}^\uparrow -
k_{\rm F}^\downarrow ) x\right]\over{( k_{\rm F}^\uparrow - k_{\rm
F}^\downarrow) x } }.
\label{eq:asymp}
\end{eqnarray}
To understand this result, we note (see the top panel
of Fig.~\ref{fig:fe_slice})
that the electron  states with wave vectors in an interval
$\delta q$ near $q=0$ (which share the value $\Delta k\simeq k_{\rm
F}$) play a special role. 
These states precess together (coherently) with spatial frequency
$k_{\rm F}^{\uparrow}-k_{\rm F}^\downarrow$. Slow dephasing begins
only after a distance $x$ where $x\delta q \sim 1$. 


The oscillatory, algebraic decay exhibited by (\ref{eq:asymp})
contrasts markedly with the
assumption of monotonic, exponential decay made by others.
\cite{Berger:1996,HZE:2001,Simanek:2001} Of course, incoherent
scattering processes may be expected to superimpose an exponential
decay on the algebraic decay we find.  The solid curves in
Fig.~\ref{fig:decay_fe} illustrate the behavior of the transmitted
spin current (\ref{xdecay}) for three free-electron models. The dashed
curves show the asymptotic behavior from (\ref{eq:asymp}). The top
panel corresponds to Fig.~\ref{fig:fe_slice} where the Fermi sphere of
the non-magnet is significantly smaller than the majority sphere and
significantly larger than the minority sphere.
The middle panel is a situation where the Fermi sphere of the
non-magnet is identical in size to the majority sphere and both are
significantly larger than the minority sphere.  Finally, the lower
panel shows results for majority and minority spheres which are,
respectively, slightly larger and slightly smaller than the Fermi
sphere of the non-magnet. This corresponds to the so-called ``$s$-$d$
model'' where the conduction electrons bands in the ferromagnet are
regarded as slightly split by exchange with localized moments.

The interfacial ``spin-filter'' makes each solid curve in
Fig.~\ref{fig:decay_fe} differ from unity already at $x=0$. The filter
is most effective when the Fermi surface of the non-magnet is poorly
matched with one or both of the Fermi surfaces of the
ferromagnet. Owing to (\ref{T}), this is consistent with our earlier
discussion where we identified the transmission probability condition
${\rm T}^\uparrow(q)\neq {\rm T}^\downarrow(q)$ as a prerequisite to the action of
the spin filter.  The subsequent decay of each curve in
Fig.~\ref{fig:decay_fe} to zero reflects the distribution of spatial
precession frequencies as we have indicated.  We have repeated these
calculations assuming that the distribution function $g_{\rm p}$ is
proportional to the velocity term in (\ref{distribution}) alone rather
than the constant term in (\ref{distribution}) alone. We find no
significant changes from the results of Fig.~\ref{fig:decay_fe}.

\begin{figure} 
\includegraphics{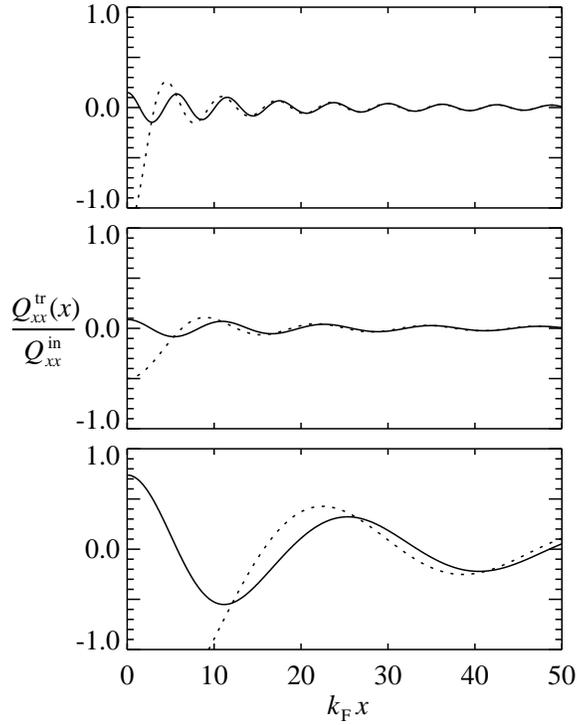} 
\caption{Decay of transverse transmitted spin current as a function of distance
from the interface for three free electron models.
In each panel, the solid curve is the exact result (\protect\ref{xdecay}) and the dashed
curve is the asymptotic result (\protect\ref{eq:asymp}). 
Top panel:
the mismatch is very large between the sizes of the
magnetic and non-magnetic 
Fermi surfaces; 
$k_{{\rm F}\uparrow}/k_{\rm F}=1.5$ and $k_{{\rm F}\downarrow}/k_{\rm F}=0.5$.
This is the model used in Fig.~\ref{fig:fe_slice}.
Middle panel:
the Fermi surfaces are identical for the non-magnet and the majority
electrons in the magnet;
$k_{{\rm F}\uparrow}/k_{\rm F}=1.0$ and $k_{{\rm F}\downarrow}/k_{\rm F}=0.5$.
Bottom panel: an $s$-$d$-like model where the mismatch is very
small between the sizes of the magnetic and non-magnetic Fermi surfaces:
$k_{{\rm F}\uparrow}/k_{\rm F}=1.1$\
and $k_{{\rm F}\downarrow}/k_{\rm F}=0.9$.}  
\label{fig:decay_fe} 
\end{figure}

We are now ready to use our computed results to find the spin-transfer
torque (\ref{interfacetorque}) for free electron models. The top and
bottom panels of Figure~\ref{fig:fe_vector} show the transverse spin
space vectors $\bf N$, ${\bf Q}^{\rm in}\cdot \hat{\bf x}$, ${\bf
Q}^{\rm ref}\cdot \hat{\bf x}$, and ${\bf Q}^{\rm tr}\cdot \hat{\bf
x}$ for the Fermi surfaces used in the top and bottom panels of
Fig.~\ref{fig:decay_fe}. We have suppressed the contraction with
$\hat{\bf x}$ in the spin current labels for clarity.  In fact, the
vectors for ${\bf Q}^{\rm ref}$ and ${\bf Q}^{\rm tr}$ represent these
quantities just at the interface.  Therefore, the reflected piece
includes the dephasing effects of differential spin rotation whereas
the transmitted piece does {\em not} include the dephasing effects of
differential spin precession. As we have seen, the latter reduces the
transmitted spin torque to zero not far from the interface.
Therefore, we have drawn the torque vector (for a unit area of
interface) so ${\bf N}={\bf Q}^{\rm in}+{\bf Q}^{\rm ref}$.  The top
panel of Fig.~\ref{fig:fe_vector} (large Fermi surface mismatch) shows
a significant dephasing of the reflected spin current. The bottom
panel of Fig.~\ref{fig:fe_vector} (small Fermi surface mismatch) shows
nearly zero reflected spin current. The reflected spin current is
exactly zero for the model (not shown) used in the middle panel of
Fig.~\ref{fig:decay_fe}.  These results show that, unless the Fermi
surface mismatch is very small, the interface effectively absorbs the
entire transverse component of incident spin current. This abrupt
change in angular momentum is the source of current-induced
spin-transfer torque at the interface between a ferromagnet and a
non-magnet.


The dashed arc labeled ${\rm Q}^{\rm sf}$ in each panel of
Fig.~\ref{fig:fe_vector} is a portion of a circle whose center is the
``tail'' position for all three spin current vectors. The radius of
this circle, compared to the length of the vector ${\bf Q}^{\rm in}$,
gives an indication of the magnitude of the spin filter
effect. Quantitatively, the circle radius is proportional to
\begin{eqnarray}
{ Q_{x}^{\rm sf} \over Q_{xx}^{\rm in} }
&=&   { 2 \over k_{\rm F}^2}
\int\limits_0^{k_{\rm F}} d q q 
\left|{R_\uparrow^*(q)}
R_{\downarrow}(q) \right|
+ { 2 \over k_{\rm F}^2}
\int\limits_0^{k_{\rm F}^\downarrow} d q q 
{ k^{\downarrow}_x(q) + k^{\uparrow}_x(q) \over  2 | k_x(q) | }
\left| T_\uparrow(q)
T_{\downarrow}(q) \right|.
\end{eqnarray}
With this definition, ${\rm Q^{sf}}$ measures the magnitude of the
total outgoing spin current (reflected plus transmitted) without
taking phase cancellation into account.  This scalar measure of the
spin filter is truly meaningful only when the reflection and
transmission amplitudes are both real and positive, which is not the
case. Nevertheless, the dashed arcs give some insight into the
efficacy of the spin filter mechanism for different free-electron
Fermi surfaces.

The foregoing makes clear that free electron models are useful for
building intuition about spin currents and spin-transfer
torque. However, there is no substitute for first-principles
calculations if we are interested in specific material interfaces. At
the very least, such calculations can be used to judge the correctness
of approximate constructs such as the $s$-$d$ model.

\begin{figure} 
\includegraphics{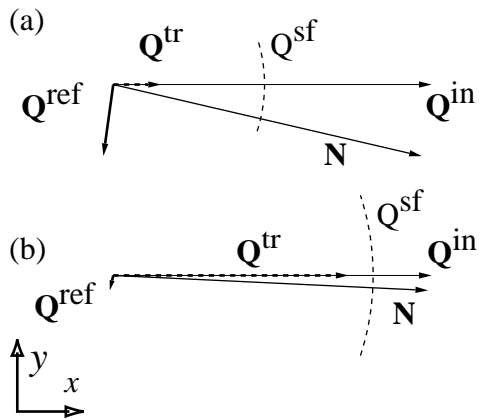} 
\caption{Graphical representation of the interfacial torque and transverse spin currents
for two free electron models. The $x$-components
are horizontal and $y$-components are vertical.  The horizontal arrow is 
the incident spin current directed along the $x$-direction.  The dashed arc
indicates the reduction in  spin current due to the
``spin-filter'' effect. The thick arrow is the reflected spin
current at $x=0$. The dashed arrow is the transmitted spin current at 
$x=0$.   The thin arrow is the final torque taking account
of the fact that precessional averaging in the ferromagnet drives 
${\bf Q}^{\rm tr}\rightarrow 0$ after a few lattice constants.  Panel (a) is
the large Fermi surface mismatch model of Fig.~\protect\ref{fig:fe_slice}. Panel (b) is  
the $s$-$d$ model of the bottom panel of Fig.~\protect\ref{fig:decay_fe}.}
\label{fig:fe_vector} 
\end{figure}

\section{Real interfaces}

In this section we repeat the calculations of Section~\ref{sec:free_dist}
for ten lattice-matched interfaces between a non-magnet and a
ferromagnet using a more realistic model of the electronic structure
for both. Specifically, we calculate the transmission and reflection
amplitudes using a linearized-augmented-plane-wave implementation of
the local-spin-density approximation. The details can be found in
Ref.~\onlinecite{Stiles:96a} and Ref.~\onlinecite{Stiles:96b}.
Compared to that earlier work, the calculations reported here use a
mesh in reciprocal space that is a factor of two denser in each
direction. For one case (Co/Cu), we checked that no changes in
relative spin currents greater than $10^{-3}$ occurred when the mesh
was made another $2\times 2$ denser.  Evanescent states (which decay
exponentially away from the interface) play a crucial role in the
calculation of the reflection and transmission amplitudes.  We have
ignored them in our spin currents computations. Their effect is to
change the wave functions in the immediate vicinity (a few atomic
layers) of the interface in such a way that there is no true
discontinuity in the transverse spin current at the interface.  As a
practical matter, this means only that the ``interfacial'' torque we
compute is--in reality--spread out over a few atomic spacings.  

The two middle panels of Fig.~\ref{fig:fs_slice_111} show a slice
through the Fermi surface of copper and the same slice through the
majority (dashed lines) and minority (solid) Fermi surfaces of cobalt
for the Co/Cu(111) system. The Fermi surface topologies here are much
more complicated than the corresponding free electron topologies
(cf. Fig.~\ref{fig:fe_slice}).  Moreover, as the Co minority Fermi
surface shows, there can be more than one pair of states for each
parallel wave vector. Consequently, we supplement every integral over
parallel wave vectors with a sum over all possible states that move
toward the interface for each parallel wave vector.  We index these
states by $n$, refer to them as associated with the $n^{\rm th}$ sheet
of the Fermi surface, and adopt the notation ${\bf k}_{n\sigma}=\{{\bf
q},k^x_{n\sigma}\}$ to label Fermi surface wave functions.  We drop
the spin index $\sigma$ in the non-magnet.

\begin{figure} 
\includegraphics[height=6in]{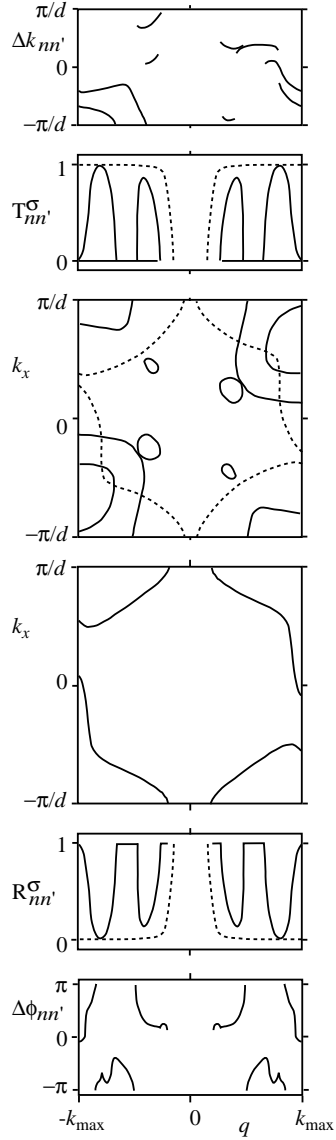} 
\caption{Same as Fig.~\protect\ref{fig:fe_slice} for a real
material interface: Co/Cu(111) with
parallel wave vectors in the $[1\overline{1}0]$
direction. The two
middle panels show the Fermi surface for the non-magnet and the
superimposed Fermi surfaces of the
majority (dashed) and minority (solid) states of the ferromagnet.
The panel just above the magnetic
Fermi surfaces is the probability for transmission into the ferromagnet
for  majority (dashed) and minority (solid) electrons. The panel just below
the non-magnetic Fermi surface is the probability for reflection back into the
non-magnet for majority (dashed) and minority (solid) electrons. The bottom
panel shows the phase in (\protect\ref{eq:deltaphi_gen}) acquired
by an electron because its spin rotates upon
reflection. The top panel shows the wave vector difference
(\protect\ref{eq:deltak_gen})
for a transmitted electron.}
\label{fig:fs_slice_111}
\end{figure}

The transverse pieces of the incident spin current
for a real interface are 
\begin{eqnarray}
Q^{\rm in}_{xx}  &=&{ 1 \over 4\pi}
\int\limits { d^2 q \over (2\pi)^2} 
\sum_n
g_{\rm p}({\bf k}_n) \sin\theta({\bf k}_n) \cos\phi({\bf k}_n)
\label{eq:qinxx_gen}
\end{eqnarray}
and
\begin{eqnarray}
Q^{\rm in}_{yx}  &=&{ 1 \over 4\pi}
\int\limits { d^2 q \over (2\pi)^2} 
\sum_n
g_{\rm p}({\bf k}_n) \sin\theta({\bf k}_n) \sin\phi({\bf k}_n).
\label{eq:qinyx_gen}
\end{eqnarray}
These differ from the corresponding free electron formulae by the
sum over the sheet index $n$. That sum is restricted to the sheets
of the Fermi surface where the electrons move {\em toward} the interface.

As before, the efficacy of the spin filter can be judged from
the interface transmission and reflection probabilities. These  
state-to-state ($n \rightarrow n'$) quantities are 
\begin{eqnarray}
{\rm T}_{ nn'}^\sigma= { v^\sigma_{n'} \over v_n }
\left|T_{\sigma nn'} \right|^2
\end{eqnarray}
and
\begin{eqnarray}
{\rm R}_{ nn'}^\sigma= {| v_{n'}| \over v_n } \left|R_{\sigma nn'} \right|^2.
\label{absv}
\end{eqnarray}
The absolute value is needed in (\ref{absv}) because $v_{n'}<0$ and
${\rm R}_{ nn'}^\sigma$
must be non-negative. Fig.~\ref{fig:fs_slice_111} shows the transmission and
reflection probabilities for 
one slice through the  Co/Cu(111)  Fermi surfaces. 

The transverse components of the reflected spin current are
\begin{eqnarray}
Q^{\rm ref}_{xx}
&=&  -{1\over 4\pi}
\int\limits { d^2 q \over (2\pi)^2} 
\sum_n 
g_{\rm p}({\bf k}_n) \sin\theta({\bf k}_n)
\sum_{n'}  
{ |v_x({\bf k}_{n'})| 
 \over | v_x({\bf k}_n) | }
{\rm Re}\left[  
R_{\uparrow n n' }^* R_{\downarrow n n'} e^{-i\phi({\bf k}_n)} 
\right]
\label{eq:qrefxx_gen}
\end{eqnarray}
and
\begin{eqnarray}
Q^{\rm ref}_{yx}
&=& - {1\over 4\pi}
\int\limits { d^2 q \over (2\pi)^2} 
\sum_n 
g_{\rm p}({\bf k}_n) \sin\theta({\bf k}_n)
\sum_{n'}  
{  |v_x({\bf k}_{n'})| 
 \over | v_x({\bf k}_n) | }
{\rm Im}\left[  
R_{\uparrow n n' }^* R_{\downarrow n n'} e^{-i\phi({\bf k}_n)} 
\right].
\label{eq:qrefyx_gen}
\end{eqnarray}
Here, the sum over $n'$ is restricted to the sheets of the Fermi
surface where the electrons move {\em away} from the interface.
Similar to the free electron case, the dephasing of the reflected
transverse spin current is determined by reflection phases $\Delta
\phi^{\rm R}_{nn'}({\bf q})$ where
\begin{eqnarray}
R_{\uparrow nn'}^* R_{\downarrow nn'} = \left|
R_{\uparrow nn'}^* R_{\downarrow nn'} \right|e^{i\Delta \phi^{\rm
R}_{nn'}({\bf q})} .
\label{eq:deltaphi_gen}
\end{eqnarray}
The bottom panel of Fig.~\ref{fig:fs_slice_111} shows that the
Co/Cu(111) phases are both more complicated and exhibit greater
dispersion than the corresponding free electron results plotted in
Fig.~\ref{fig:fe_slice}.  

The transverse pieces of ${\bf Q}^{\rm in}$
and ${\bf Q}^{\rm ref}$ written above are closely related to the {\em
mixing conductance}, $G_{\rm mix}$, introduced by Brataas {\em et al.}
\cite{Brataas:2001} and computed recently by Xia {\em et al.}
\cite{Xia} In our notation,
\begin{eqnarray}
G_{\rm mix}={e^2 \over h}A
\int\limits{d^2 q \over (2\pi)^2}
\sum_n
\left[
1-
\sum_{n'}
{|v_x({\bf k}_{n'})|
\over|v_x({\bf k}_n)|}
R_{\uparrow n n'}^*R_{\downarrow n n'}
\right].
\label{eq:gmix}
\end{eqnarray}
This formula is relevant to situations where $g_{\rm p}({\bf k}_n)$,
$\theta({\bf k}_n)$, and $\phi({\bf k}_n)$ in
(\ref{eq:qinxx_gen})-(\ref{eq:qrefyx_gen}) are all constants--a
restriction implicit in the Landauer description of transport. In that
case, the real and imaginary parts of $G_{\rm mix}$ are proportional
to the $xx$ and $yx$ components of ${\bf Q}^{\rm in}+ {\bf Q}^{\rm
ref}$. From (\ref{interfacetorque}), the latter is proportional to the
spin-transfer torque if we neglect the transverse part of the
transmitted spin current.\cite{Brataas:2001,us:2001}  For the systems
treated by both of us, our numerical results for spin-transfer torque
agree semi-quantitatively with the mixing conductance calculations of
Xia at al.

The transverse transmitted spin currents are
\begin{equation}
Q^{\rm tr}_{xx}(x) =  { 1\over 4\pi }\int\limits { d^2 q \over
(2\pi)^2} \sum_n g_{\rm p}({\bf k}_n) { \sin\theta({\bf k}_n) \over |
v_x({\bf k}_n) | }{\rm Re}\left[ e^{-i\phi({\bf k}_n)} \sum_{n'',n'}
T_{\uparrow n n''}^*   T_{\downarrow n n'}\Phi_{n'' n'} ({\bf
q},x)e^{i ( k^x_{n'' \uparrow} - k^x_{n' \downarrow} ) x }\right]
\label{eq:qzx_gen}
\end{equation}
and
\begin{equation}
Q^{\rm tr}_{yx}(x) =  { 1\over 4\pi }
\int\limits { d^2 q \over (2\pi)^2} 
\sum_n 
g_{\rm p}({\bf k}_n) 
{ \sin\theta({\bf k}_n) \over | v_x({\bf k}_n) | }
{\rm Im}\left[ e^{-i\phi({\bf k}_n)} 
\sum_{n'',n'} 
T_{\uparrow n n''}^*
   T_{\downarrow n n'}
\Phi_{n'' n'} ({\bf q},x)
e^{-i ( k^x_{n'' \uparrow} - k^x_{n' \downarrow} ) x }
\right].
\label{eq:qzy_gen}
\end{equation}
Apart from the sums over $n'$ and $n''$ (both restricted to sheets of
the ferromagnetic Fermi surfaces where electrons move away from the
interface), these formula are less simple than the corresponding free
electron results (\ref{xdecay}) and (\ref{ydecay}) for two reasons.
First, the transmission amplitudes, $T_{\sigma n n'}$, are complex
rather than real. Second, the Bloch wave functions $\psi_\sigma({\bf
R},x,{\bf k}_{n \sigma})$ have a non-trivial dependence on the spatial
variable ${\bf R}$ parallel to the interface plane. For the latter reason,
the transmitted spin currents contain a factor $\Phi_{n'' n'} ({\bf
q},x)$ defined by
\begin{eqnarray}
\Phi_{n n'} ({\bf q}, x) e^{-i ( k^x_{n \uparrow} - k^x_{n'
\downarrow} ) x }&=& {1\over 2A}\int\limits_{v_x>0} d{\bf R}
\left[\psi_{\uparrow}^*({\bf r},{\bf k}_{n
\uparrow})\hat{v}_x\psi_{\downarrow}  ({\bf r},{\bf k}_{n'\downarrow})
-\psi_{\downarrow}  ({\bf r},{\bf
k}_{n'\downarrow})\hat{v}_x\psi_{\uparrow}^*({\bf r},{\bf k}_{n
\uparrow})\right] .
\label{eq:phase}
\end{eqnarray}
This yields 
\begin{eqnarray}
\Phi_{n n'} ({\bf q}, x) 
&=& {v_x^{n' \downarrow} +
v_x^{n' \downarrow}\over 2}
\end{eqnarray}
when free electron wave functions are used in (\ref{eq:phase}).
Otherwise, $\Phi_{n'' n'} ({\bf q},x)$ is a complex, periodic function
of $x$ with period equal to one layer spacing. Thus, it can be
calculated once and propagated from layer to layer.  A related factor
enters the reflected spin currents (\ref{eq:qzx_gen}) and
(\ref{eq:qzy_gen}). However, because the spin up and spin down wave
functions are the same in the non-magnet, it reduces to the velocity
factor in the numerator of those expressions.

Given the foregoing, it is sensible to define transmission phases
$\Delta \phi^{\rm T}_{nn'}({\bf q})$ so that
\begin{eqnarray}
T_{\uparrow nn'}^* T_{\downarrow nn'}\Phi_{n'' n'} ({\bf q},x=0)  = 
\left| T_{\uparrow nn'}^* T_{\downarrow nn'} \Phi_{n'' n'} ({\bf q},x=0) \right|
e^{i\Delta \phi^{\rm T}_{nn'}({\bf q})}.
\label{eq:deltaphi_genT}
\end{eqnarray}
This tells us that, unlike free electrons, the spins of Bloch
electrons generally rotate when they transmit through a real material
interface.  If the distribution of transmission phases is broad,
substantial cancellation of the transmitted spin current occurs at
$x=0$ when we sum over all transmitted electrons. This effect is
independent of the spin filter, which also acts at $x=0$.


Any spin current that survives to propagate into the ferromagnet
rapidly disappears due to differential spatial precession. The
(generalized) spatial precession frequency is determined by the
difference in wave vector for different sheets of the majority and
minority Fermi surfaces:
\begin{eqnarray}
\Delta k_{n'n''}= k^x_{n' \downarrow} - k^x_{n'' \uparrow}.
\label{eq:deltak_gen}
\end{eqnarray}
The top panel of Fig.~\ref{fig:fs_slice_111} illustrates the
distribution of $\Delta k_{n'n''}$ for a Fermi surface slice of
Co/Cu(111).  The large dispersion seen there suggests that the spin
current decays very quickly in the ferromagnet. This is confirmed by
Fig.~\ref{fig:decay}, which shows the computed decay of the transverse
spin current for three interface orientations of Co/Cu. The non-zero
value of the dashed curves at $x=0$ shows that a large amount of rotation
occurs upon transmission.  The Fermi surfaces are more complicated than
the free electron models, so the initial decay is more complicated
also.  Nevertheless, both the (111) and (110) orientations settle into
behavior that is readily characterized as a damped precession.  The
amplitude of the precession for the (100) orientation is so small that
it is difficult to see whether it is precessing or not.  In
general, there could be several decaying precessions with different
precession rates and different amplitudes.

It is worth noting that none of these curves (or analogous curves for
the other material pairs we have studied) resembles the the bottom
panel of Fig.~\ref{fig:decay_fe} appropriate to the $s$-$d$ model.
This lack of agreement is consistent with the fact that essentially
{\em all} the Fermi surface wave functions in third-row ferromagnets
contain as much ``localized'' $3d$ character as ``delocalized'' $4s$
character. \cite{unpublished}

\begin{figure} 
\includegraphics{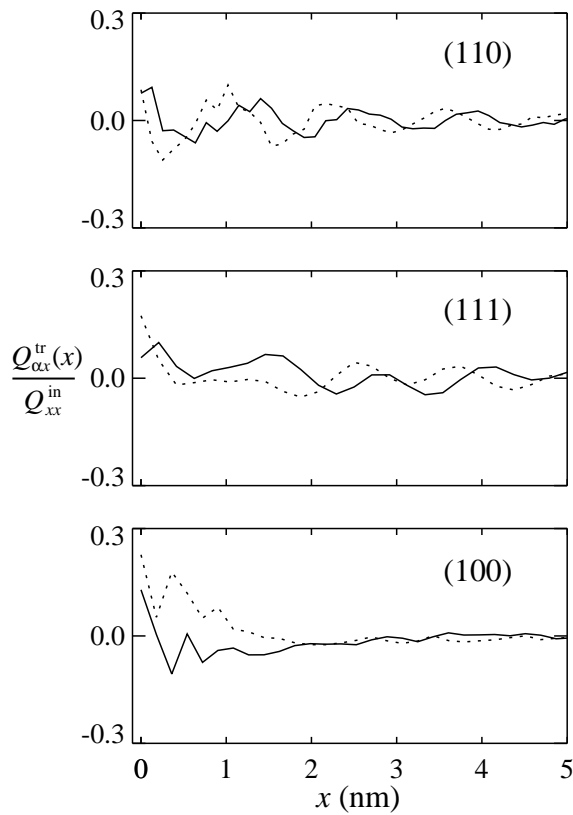} 
\caption{Decay of transverse transmitted spin current as a function
of distance from the interface for three orientations of Co/Cu. For a unit
incident transverse polarization, the solid curve in each panel
is $Q_{xx}(x)$. The dashed curve in each panel is $Q_{yx}(x)$.}
\label{fig:decay} 
\end{figure}

Figure~\ref{fig:vector} graphically summarizes our first-principles
spin current calculations for ten different interface combinations.
The vectors labeled ${\bf Q}^{\rm ref}$ and ${\bf Q}^{\rm tr}$
correspond to $x=0$ and reflect the effect of spin filtering and spin
rotation only. ${\bf Q}^{\rm ref}$ is very small and, as we have
emphasized, ${\bf Q}^{\rm tr}\rightarrow 0$ after a few lattice
constants. Therefore, the torque per unit area of interface is ${\bf
Q}^{\rm in}+{\bf Q}^{\rm ref} \simeq {\bf Q}^{\rm in}$.  Due to spin
filtering, differential spin rotation, and differential precession,
nearly all of the incident transverse spin current is absorbed in the
immediate vicinity of the interface.  For Co/Cu, Fe/Ag, and Fe/Au, the
spin-filter accounts for somewhat more than half of the effect and the
interface dephasing for the rest.  For Ni/Cu and Fe/Cr the spin filter
effect is weaker.  For Ni/Cu, the decay of the precessing transmitted
spin current plays a large role.

Of course, our calculations pertain to ideal, lattice-matched
interfaces.  A variety of effects make the interfacial absorption of
transverse spin even more efficient. We have mentioned already that
scattering in the ferromagnetic layer increases the rate of decay of
the precession.  Steps at the interface lead to increased dephasing
for both reflection and transmission.  For thin layers where the decay
of the precession might not be complete, the dephasing on passing
through the second interface generally leads to a further decay of the
transverse spin current.  Thickness fluctuations further reduce the
spin current.

\begin{figure*} 
\includegraphics{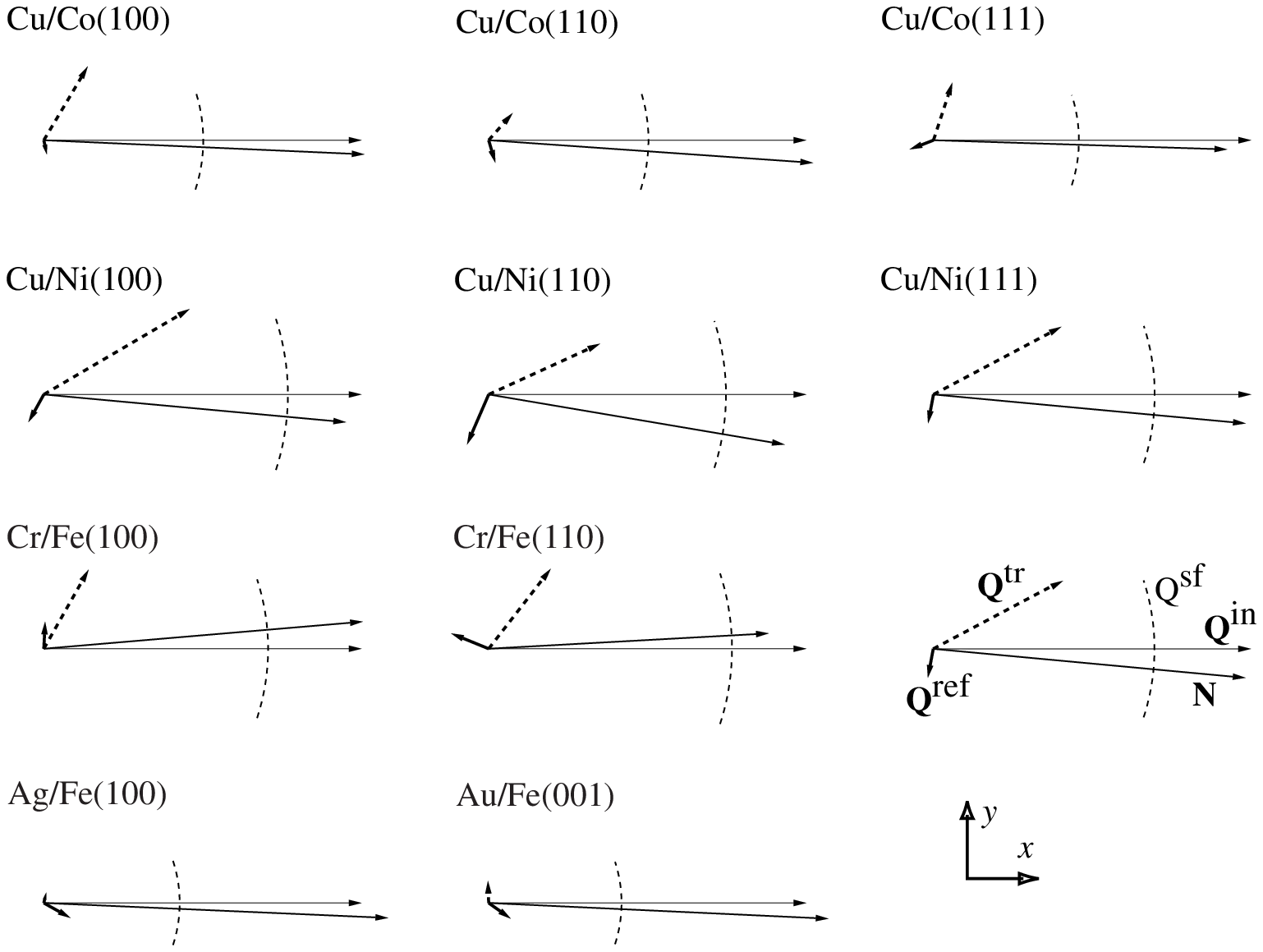} 
\caption{Graphical representation of the interfacial torque and transverse
spin currents for a series of real interfaces.
The $x$-components
are horizontal and $y$-components are vertical.  The horizontal arrow is 
the incident spin current directed along the $x$-direction.  The dashed arc
indicates the reduction in  spin current due to the
``spin-filter'' effect.  The thick arrow is the reflected spin
current at $x=0$. The dashed arrow is the transmitted spin current at 
$x=0$.  The thin arrow is the final torque taking account
of the fact that precessional averaging in the ferromagnet drives 
${\bf Q}^{\rm tr}\rightarrow 0$ after a few lattice constants.}  
\label{fig:vector} 
\end{figure*}

The interface torque we compute is interesting because of the recent
demonstrations of current-induced magnetization
switching. \cite{Myers:1999,Grollier:2001} However, the material pair
that optimizes the switching is not determined by the conversion of
the spin current into a torque.  This process is the same for all of
the interfaces considered.  For the ideal interfaces considered here,
the optimum choice depends on the ability of the material pair to
generate a spin current in the first place.  This depends on the
spin-dependent interface resistances and the spin-dependent bulk
conductivities.  The Fe/Au and Fe/Ag pairs have the strongest spin
dependence of the interface resistance.\cite{Stiles:2000} However, in
reality the optimum combination will likely depend on growth
considerations.  The general mismatch between the body-centered-cubic
Fe lattice and the face-centered-cubic Au or Ag lattice will probably
lead to poor growth, unless the interface is forced to be (100)
(where the rotated lattices match quite well) and the number of steps at
the interface is kept quite small.

\section{Summary}

In this paper, we used free electron models and first principles
electronic structure calculations to study the
spin-transfer torque that occurs when a
spin-polarized current flows from a non-magnet into a ferromagnet
through a perfect interface.  The origin of the torque is a transfer
of spin angular momentum from the conduction electrons to the
magnetization of the ferromagnet. The origin of the angular momentum
transfer is the absorption of transverse spin current by the 
interface. We identified three
distinct processes that contribute to the absorption: (1) spin-dependent
reflection and transmission; (2) rotation of reflected and transmitted
spins; and (3) spatial precession of spins in the ferromagnet.  When
summed over all Fermi surface electrons, these processes reduce the
transverse component of the transmitted and reflected spin currents to
nearly zero for most systems of interest. Therefore, to a good
approximation, the torque on the magnetization is proportional to the
transverse piece of the incoming spin current.

To be more quantitative, we used the 
analogy between charge current and spin current
to show that a spin current flowing in the $+\hat{\bf x}$ direction
(perpendicular to the interface) delivers a torque per unit area
\begin{equation}
{{\bf N}_{\rm c} \over A}=\,({\bf Q}^{\rm in}-{\bf Q}^{\rm tr}
+{\bf Q}^{\rm ref})\cdot \hat{\bf x}
\label{lastorque}
\end{equation}
to a microscopically small region around the interface.
Here,  ${\bf Q}^{\rm in}$, ${\bf
Q}^{\rm tr}$, and ${\bf Q}^{\rm ref}$ are the incident, transmitted, and
reflected spin currents computed using incident, transmitted, and
reflected wave functions. We found the latter by solving the
one-electron stationary-state scattering problem. In
the quasi-classical approximation, the total spin
current is the sum of contributions from every conduction electron.

Quite generally, the component of ${\bf N}_{\rm c}$
parallel to the ferromagnetic magnetization is zero. This is
consistent with our classical intuition.  On the other hand, we found
that the transverse components of ${\bf Q}^{\rm tr}$ and ${\bf Q}^{\rm
ref}$ relevant to (\ref{lastorque}) are also zero (or nearly so),
except in very exceptional
cases. This means that the entire transverse spin current is absorbed
(transferred to the magnetization) in the immediate vicinity of the
interface. As indicated above, this is so due to spin filtering, 
differential spin reflection, and differential spin precession.

The spin-filter effect occurs because the wave function for an
incident electron with a non-zero spin component transverse to the
magnetization spin can always be written as a linear combination of
spin-up and spin-down components. Then, because the reflection and
transmission amplitudes differ for up and down spins, the up and down
spin content of the reflected and transmitted wave functions (which
are spatially separated) differ from both each another and from the
incident state. The spin currents directly encode this information. As
a result, the right side of (\ref{lastorque}) is non-zero. This is a
one-electron effect that operates independently for every electron.

The two other effects that help drive the transverse parts of ${\bf
Q}^{\rm tr}$ and ${\bf Q}^{\rm ref}$ to zero occur when we sum over
the entire ensemble of conduction electrons. The first arises because
the spin of an electron generally rotates when it is reflected or
transmitted at the interface between a non-magnet and a
ferromagnet. The rotation is non-classical and the amount of rotation
differs considerably for electrons with wave vectors from different
portions of the Fermi surface.  Phase cancellation occurs when we sum
over all electrons. In the end, we find that very little remains of
the reflected transverse spin current. The cancellation of the
transmitted spin current is less dramatic.

Finally, due to exchange splitting, the electrons that
transmit into the ferromagnet possess spin-up and spin-down components
with the same total energy, ${\rm E_F}$, but different kinetic energy
and so different wave vectors.  This
implies that each electron spin precesses (in space) as it propagates
away from the interface.  However, like the spin rotation angles, the
spatial precession frequency varies considerably over the Fermi
surface. Consequently, rapid dephasing of the transverse spin
components of the individual electrons occurs as the conduction
electron ensemble propagates into the ferromagnet. The net result is a
precessing spin current that damps out algebraically within a few
lattice constants of the interface.

Our first principles calculations show that the relative importance of
these three mechanisms differs for different materials pairs and also
for different crystallographic orientations for the same material
pair. Nevertheless, the final result is the same in all cases: the
transverse spin current essentially disappears at the interface.  The concomitant
transfer of angular momentum delivers a torque to the magnetization in
the immediate vicinity of the interface.

\section{Acknowledgment}

One of us (A.~Z.) gratefully acknowledges support for this research
from the National Science
Foundation under grant DMR-9820230.


\begin{thebibliography}{24}

\bibitem{Slonczewski:1996}
J. C. Slonczewski, J. Magn. Magn. Mater. {\bf 159}, L1 (1996).

\bibitem{Berger:1996}  
L. Berger, Phys. Rev. B {\bf 54}, 9353 (1996).

\bibitem{others}
  M. Tsoi, A. G. M. Jansen, J. Bass, W. C. Chiang, M. Seck, V. Tsoi, 
   P. Wyder, Phys. Rev. Lett. {\bf 80}, 4281 (1998);
  J. Z. Sun, J. Magn. Magn. Mater. {\bf 202}, 157 (1999);
J.-E. Wegrowe, D. Kelly, Ph. Guittienne, Y. Jaccard, and J.-Ph. Ansermet,
Europhys. Lett. {\bf 45}, 626 (1999).

\bibitem{FertBruno:1994} 
A. Fert and P. Bruno, in {\it Ultrathin Magnetic
Structure II}, edited by B. Heinrich and J.A.C. Bland (Springer-Verlag, Berlin,
1994).

\bibitem{Myers:1999}  
E. B. Myers,    D. C. Ralph,    J. A. Katine,
R. N. Louie,    R. A. Buhrman, Science {\bf 285}, 867
(1999);
J. A. Katine, F. J. Albert, R. A. Buhrman, E. B. Myers,
D. C. Ralph, Phys. Rev. Lett. {\bf 84}, 3149 (2000).

\bibitem{Grollier:2001}
J. Grollier, V. Cros, A. Hamzic, J. M. George,
H. Jaffr\'es, A. Fert, G. Faini, J. Ben Youssef,
H. Legall, Appl. Phys. Lett. {\bf 78}, 3663 (2001).

\bibitem{Wegrowe:2001}
J.-E. Wegrowe, D. Kelly, T. Truong, Ph. Guittienne, and J.-Ph. Ansermet,
Europhys. Lett. {\bf 56}, 748 (2001).

\bibitem{Weber:2001}
W. Weber, S. Rieseen, and H. C. Siegmann, Science {\bf 291}, 1015 (2001).

\bibitem{Urban:2001}
R. Urban, G. Woltersdorf, and B. Heinrich, Phys. Rev. Lett. {\bf 87}, 217204 (2001).

\bibitem{Prinz:1999}
G. A. Prinz, J. Mag. Mag. Mat. {\bf 200}, 57 (1999).


\bibitem{Baz:2000}
Ya. B. Bazaliy, B. A. Jones, and S.-C. Zhang, Phys. Rev. B {\bf 57},
R3213 (1998); Ya. B. Bazaliy, B.A. Jones, and S.-C. Zhang, J. Appl. Phys. 
{\bf 89}, 6793 (2001); Ya. B. Bazaliy, B. A. Jones, and S.-C. Zhang,
{\it Current-induced magnetization switching in small domains of different
anisotropies}, cond-mat/0009034.

\bibitem{Sun:2000}
J. Z. Sun, Phys. Rev. B {\bf 62}, 570 (2000).

\bibitem{HZE:2001}
C. Heide, P. E. Zilberman, and R. J. Elliott, Phys. Rev. B {\bf 63}, 64424 (2001);
C. Heide, Phys. Rev. Lett. {\bf 87}, 197201 (2001); C. Heide, Phys. Rev. B {\bf 65},
054401 (2001).


\bibitem{Miltat:2001}
J. Miltat, G. Albuquerque, A. Thiaville, and C. Vouille, J. Appl. Phys. {\bf 89}, 6982
(2001).

\bibitem{ZL:2002}
S. Zhang and P. M. Levy, Phys. Rev. B {\bf 65}, 052409 (2002).

\bibitem{Tserk:2001}
Y. Tserkovnyak, A. Brataas, and G.E.W. Bauer,
{\it Enhanced Gilbert damping in Thin Ferromagnetic Films}, cond-mat/0110247.

\bibitem{spinaccumulation} 
A. G. Aronov, JETP Lett. {\bf 24}, 32 (1977);
M. Johnson and R. H. Silsbee, Phys. Rev. B {\bf 35},
4959 (1987); P. C. van Son, H. van Kempen, and P. Wyder, Phys. Rev. Lett. {\bf 58},
2271 (1987).

\bibitem{Weg:2000}
J.-E. Wegrowe, Phys. Rev. B {\bf 62}, 1067 (2000).

\bibitem{Brataas:2001}
A. Brataas, Yu. V. Nazarov, and G. E. W. Bauer,
Phys. Rev. Lett. {\bf 84}, 2481 (2000); A. Brataas, Yu.V. Nazarov, and
G. E. W. Bauer, Eur. Phys. J. B {\bf 22}, 99 (2001).

\bibitem{Waintal:2000}  
X. Waintal,    E. B. Myers,    P. W. Brouwer,
D. C. Ralph, Phys. Rev. B {\bf 62}, 12317 (2000).

\bibitem{Hernando:2000}
D. H. Hernando, Yu.V. Nazarov, A. Brataas, and
G. E. W. Bauer, Phys. Rev. B {\bf 62}, 5700 (2000). 

\bibitem{us:2001}
M.D. Stiles and A. Zangwill, {\it Non-collinear spin transfer in
Co/Cu/Co multilayers}, cond-mat-0110275.

\bibitem{Xia}
K. Xia, P. J. Kelly, G. E. W. Bauer, A. Brataas, and I. Turek, {\it Spin Torques in 
Ferromagnetic/Normal Metal Structures}, cond-mat/0107589.

\bibitem{sign_convention}
Eq.~(13) is equivalent to Eq.~(10) in [22]. That relation was written
in terms of the magnitude of the current. Thus, the term for the
reflected spin flux appears there with no minus relative to the
transmitted spin flux because the reflected current is in the opposite
direction to the transmitted current.


\bibitem{Gottfried} 
We exploit the fact that the cross section for a time-dependent
wave-packet scattering process is determined by the scattering
amplitude (here, the transmission and reflection amplitudes) of the
corresponding stationary-state, wave-function matching problem. See,
{\em e.g.}, K. Gottfried, {\em Quantum Mechanics} (Addison-Wesley,
Reading, MA, 1966), Section~12.

\bibitem{precess} 
Spatial precession occurs when a coherent
superposition of up and down spin states have the same energy ($E_{\rm
F}$ in this case) but different wave vectors.  The more familiar case
of temporal precession occurs when a coherent superposition of up and
down spin states have the same wave vector but different energies. In
both cases, an accumulation of phase changes the orientation of the
spin vector.


\bibitem{Ziman}
J. M. Ziman, {\em Principles of the the theory of solids}, (University Press,
Cambridge, 1972), Chapter 7.

\bibitem{Stilesintegral}
 M. D. Stiles, Phys. Rev. B{\bf 48}, 7238 (1993).

\bibitem{Simanek:2001} E. Simanek, Phys. Rev. B {\bf 63}, 224412 (2001).

\bibitem{Stiles:96a} M. D. Stiles, J. Appl. Phys. {\bf 79}, 5805 (1996).

\bibitem{Stiles:96b} M. D. Stiles, Phys. Rev. B {\bf 54}, 14679 (1996).
 
\bibitem{Stiles:2000} M. D. Stiles and D. R. Penn, Phys. Rev. B {\bf 61}, 3200 (2000).

\bibitem{unpublished} M. D. Stiles, (unpublished).

\end{thebibliography}
\end{document}